\documentclass[letter,12pt]{article}
\pdfoutput=1 

\usepackage{jheppub} 

\usepackage[utf8]{inputenc}

\usepackage[russian,english]{babel}
\usepackage[T1,T2A]{fontenc} 

\usepackage{amsmath}

\usepackage{ upgreek }
\usepackage{physics}

\def\bea{\begin{eqnarray}}
\def\eea{\end{eqnarray}}

\title{Zero modes of local operators in 2d CFT on a cylinder}

\author[a,b]{Anatoly Dymarsky,} 
\author[b, c]{Kirill Pavlenko}
\author[d]{and Dmitry Solovyev} 
 
\affiliation[a]{University of Kentucky,\\Lexington, KY, USA 40506\\}
\affiliation[b]{Skolkovo Institute of Science and Technology,\\Skolkovo Innovation Center, Moscow, Russia\\}
\affiliation[c]{Moscow Institute of Physics and Technology, Dolgoprudny 141700, Russia\\}
\affiliation[d]{Saint Petersburg State University, Saint Petersburg, Russia}
\emailAdd{a.dymarsky@uky.edu}
\emailAdd{kirill.pavlenko@skoltech.ru}
\emailAdd{dimsol42@gmail.com}

\abstract{
Studies of Eigenstate Thermalization Hypothesis (ETH) in two-dimensional CFTs call for calculation of the expectation values of local operators in highly excited energy eigenstates. This can be done efficiently by representing zero modes of these operators in terms of the Virasoro algebra generators. In this paper we present a pedagogical introduction explaining how this calculation can be performed analytically or using computer algebra. We illustrate the computation of zero modes by a number of examples and list explicit expressions for all local operators from the vacuum family with the dimension of less or equal than eight. Finally, we derive an explicit expression for the quantum KdV generator  $Q_7$ in terms of the Virasoro algebra generators. The obtained results can be used for quantitative studies of ETH at finite value of central charge. }

\begin{document} 
\maketitle
\flushbottom

\section{Introduction}
\label{sec:intro}
Emergence of statistical thermodynamics from quantum mechanics is one of the exciting open questions of theoretical physics. It has became an accepted paradigm that universality of thermal equilibrium can be explained in terms of Eigenstate Thermalization \cite{deutsch1991quantum,srednicki1994chaos,rigol2008thermalization}. The latter is an expectation that local properties of individual microstates -- energy eigenstates of spatially extended sufficiently complex quantum system -- may only  depend on thermodynamically relevant quantities, i.e.~in most cases only on energy density of the microstate. Technically, this means that the diagonal matrix elements  of local observables 
\bea
\label{ethansatz}
A_{ii}\equiv \langle E_i|A|E_i\rangle=f_A(E_i),
\eea
are smooth functions of energy density $E_i/V$.

When the system, besides energy, possesses a number of additional local (or quasi-local \cite{ilievski2015quasilocal}) conserved  quantities, as is normally the case for integrable systems, densities of these additional conserved charges are also thermodynamically relevant. This simply means emerging equilibrium state will be dependent on these additional quantities, or, more accurately their densities. Further assuming local properties of individual microstates (energy eigenstates) will be dependent on these charges, but otherwise will be physically equivalent, we arrive at the notation of generalized Eigenstate Thermalization Hypothesis \cite{rigol2007relaxation,cassidy2011generalized,vidmar2016generalized}, 
\bea
\label{geth}
A_{ii}=f_A(Q_k(E_i)),
\eea
where $f_A$ is a smooth function of its arguments,  $Q_k$ are the conserved charges, and $Q_k(E_i)$ are the charge values associated with an individual energy eigenstate $|E_i\rangle$. 

Eigenstate Thermalization Hypothesis has been established numerically 
for a range of  lattice models \cite{d2016quantum}. It is therefore natural to ask if Eigenstate Thermalization  also
applies to conformal field theories\cite{de2016remarks,sonner2017eigenstate,basu2017thermality,lashkari2018eigenstate,lashkari2018universality,faulkner2018probing,lin2016thermality,basu2017thermality,Guo2018pvi,
He2017txy,GGE,GGE2, GETH,maloney2018generalized,datta2019typicality,besken2019virasoro}, which can be thought of as a continuous limit of lattice systems at criticality. This assumes conformal field theory is quantized on a cylinder of finite spatial size. The object of interest is the diagonal matrix element 
$\langle E|A|E\rangle$, where $|E\rangle$ is the eigenstate of the CFT Hamiltonian and $A$ is a local observable. 
 In case of two-dimensional theories, which is a focus of this work, there is a number of specifics which need to be taken into account. First, the theory is split into ``left'' and ``right'' non-interacting sectors, such that in conventional units the Hamiltonian $H$ is a sum of $L_0-c/24$ and ${\bar L}_0-c/24$. We can assume that $|E\rangle$ is a tensor product of two eigenstates in each sector, while $A$ is a product of local holomorphic and anti-holomorphic components.  Then the calculation factorizes and in the rest of the paper we focus on the left-moving sector only. Second, when $A$ is a global descendant, its expectation value in the eigenstate of $L_0-c/24$  is zero, thus in the following we assume $A$ is a quasi-primary. Third, and final point, we have two distinct cases to consider: when $A$ is a Virasoro descendant of identity or when $A$ is not from the vacuum family.  In two dimensions thermal expectation values of all operators outside of the vacuum family vanish (this also applies to KdV Generalized Gibbs Ensemble). Thus a prediction of ETH in $d=2$ for such operators  would be that $\langle E|A|E\rangle$ vanish when $|E\rangle$ is sufficiently heavy \cite{lashkari2018eigenstate}. This is a physically straightforward but challenging question to verify, as the behavior of individual OPE coefficients which govern this matrix element is currently outside of our theoretical control. To summarize, we are interested in  a special set of observables  -- quasi-primary operators from the vacuum family, for which matrix elements $A_{ii}$ are non-zero, universal and fixed by the Virasoro algebra. 

The spectrum of $2d$ theories is highly degenerate and therefore the choice of $|E\rangle$ is not unique.  This is related to the fact that $2d$ theories possess an infinite number of local conserved quantum KdV charges \cite{bazhanov1996integrable,bazhanov1997integrable, bazhanov1999integrable}, which must be taken into account in the context of eigenstate thermalization. In recent papers \cite{GGE,GGE2,GETH}  we have demonstrated that organizing descendant states in the eigenstates of the KdV hierarchy allows one to formulate the diagonal part of the generalized Eigenstate Thermalization Hypothesis \eqref{geth}. 

Any descendant state, including the KdV eigenstates $|E\rangle$ can be written as a linear combination of Virasoro algebra generators acting on some primary state $\Delta$,
\bea
\nonumber
|E\rangle = \sum_{k_1,\dots k_m} c_{k_1,\dots,k_m} L_{-k_1}\dots L_{-k_m}|\Delta\rangle,\qquad k_1+\dots k_m=n,\qquad E=\Delta+n,\quad k_i>0.
\eea
Even if coefficients $c_{k_1,\dots,k_m}$ are known explicitly, to calculate the diagonal matrix element $\langle E|A|E\rangle$, one needs to know an explicit form of the zero mode of the local operator $A$ in terms of the Virasoro algebra generators, 
\bea
A_0=\oint {dw\over 2\pi} A(w)\equiv {1\over 2\pi}\int_0^{2\pi} du\, A(w).
\eea
Here  $w = u + i \tau$ is a holomorphic coordinate on the cylinder, and $u$ is a periodic coordinate on a spacial circle of length $2 \pi$. The coordinate $w$ is related to the conventional coordinate on the plane by $z = e^{-i w}$.
Local operator $A$ admits the following mode expansion 
\bea
A(w)=\sum_{n=-\infty}^{\infty} A_n e^{-in w}.
\eea
Since we assumed $A$ is from the vacuum family, it is ``built'' out of energy-momentum tensor $T(w)$ whose mode expansion is
\bea
\label{T}
T(w) = -\frac{c}{24} + \sum_{n = - \infty}^{+ \infty} L_{n} e^{-i n w}.
\eea
The operators $L_n$ satisfy conventional commutation relations of Virasoro algebra
\bea
\label{virasoro}
[L_n, L_m] = (n-m) L_{n+m} + \frac{c}{12} (n^3 - n) \delta_{n+m,0}.
\eea
In practical terms $A$  are appropriately regularized polynomials in stress-energy tensor  $T(w)$ and its holomorphic derivatives. 

Among the local  (not necessarily quasi-primary) operators  there is a family of special densities which, upon integration, give rise to the tower of mutually commuting quantum KdV charges $Q_{2k-1}$. There is one such density $J_{2k}$ for each even dimension $2k$ such that 
\bea
\label{kdv}
Q_{2k-1} = \oint \frac{dw}{2 \pi} J_{2k}(w).
\eea
The general expression for $J_{2k}$ is not known. However, they are uniquely determined up to normalization by the requirement of commutativity $[Q_{2k-1}, Q_{2l -1}] = 0$ and scaling homogeneity. First few densities $J_{2k}(w)$ are
\bea
\label{dens}
J_2(w) &=& T, \\
J_4(w) &=& (TT), \\
J_6(w) &=& (T(TT)) + \frac{c+2}{12} (\partial T \partial T),\\
J_8(w)&=& (T(T(TT))) + \frac{c+2}{3}(T(\partial T \partial T)) + \frac{2c^2 -17c - 42}{360} (\partial^2 T \partial^2 T).
\eea
Here the parentheses $(\cdot)$ denote normal ordering which will be defined in the next section. 
Explicit form of $Q_{2k-1}$, i.e.~the zero modes of $J_{2k}$, in terms of the Virasoro algebra generators would allow computing the spectrum of $Q_{2k-1}$ at finite $c$ in terms of computer algebra. 

In this paper we pedagogically develop the machinery of calculating zero modes of local operators from the vacuum family and calculate explicit expressions for all quasi-primaries of the dimension less or equal than eight. We also calculate the explicit expression for $Q_7$. Some of these results, albeit in a simplified thermodynamic limit, when certain terms can be neglected, were already used by us in previous studies of ETH in \cite{GETH}.
We plan to use the explicit results obtained in this paper to further elucidate ETH in 2d CFT in the subsequent works. 

The paper is organized as follows. In section \ref{sec:zeromodes} we develop general technique for calculating  the zero modes of the normal ordered product of local operators. In section \ref{sec:q3} we show in detail how this  technique applies in the simplest case and calculate first non-trivial qKdV charge $Q_3$. In section \ref{sec:QP} we list the explicit expressions of the zero modes for all quasi-primary operators with the dimension of less or equal to eight. 
Then, in section \ref{sec:q7} we give the explicit expression for $Q_7$ in terms of the Virasoro algebra generators and verify its consistency in the large $c$ limit and also using the constraints coming from the  (9,2) minimal model. 
We conclude in section \ref{sec:con}.

\section{Zero modes of the operator product}
\label{sec:zeromodes}

In this section we will describe general technique of finding zero modes of the product of arbitrary local operators $A(w)$ and $B(w)$ on a cylinder worldsheet, namely	
\bea
\label{zerom}
(AB)_0 = \oint \frac{dw}{2\pi} (AB)(w),
\eea 
where the contour is taken over a spacial circle. The operator $A(w)$ and $B(w)$ are assumed to have analytic mode expansion 
\bea
\label{modes}
A(w) = \sum_{n = - \infty}^{+\infty} A_n e^{-inw},
\eea
and similarly for $B(w)$.
The parentheses in $(AB)(w)$ denote normal ordering of the operators and is defined as
\bea
\label{normal}
(AB)(w) = \oint_w \frac{dz}{2\pi i} \frac{1}{z -w} \mathcal{T} \left( A(z) B(w)\right),
\eea
where the integration is performed over the circle around $w$ and  the symbol $\mathcal{T}$ stands for ``chronological ordering,'' i.e.
\bea
\label{chrono}
\mathcal{T} \left( A(z) B(w)\right) = \begin{cases} A(z) B(w), & \mbox{if } \Im z < \Im w,    \\ B(w) A(z), & \mbox{if } \Im z > \Im w. \end{cases}
\eea

\begin{figure}[h]
\centering 
\includegraphics[width=.60\textwidth]{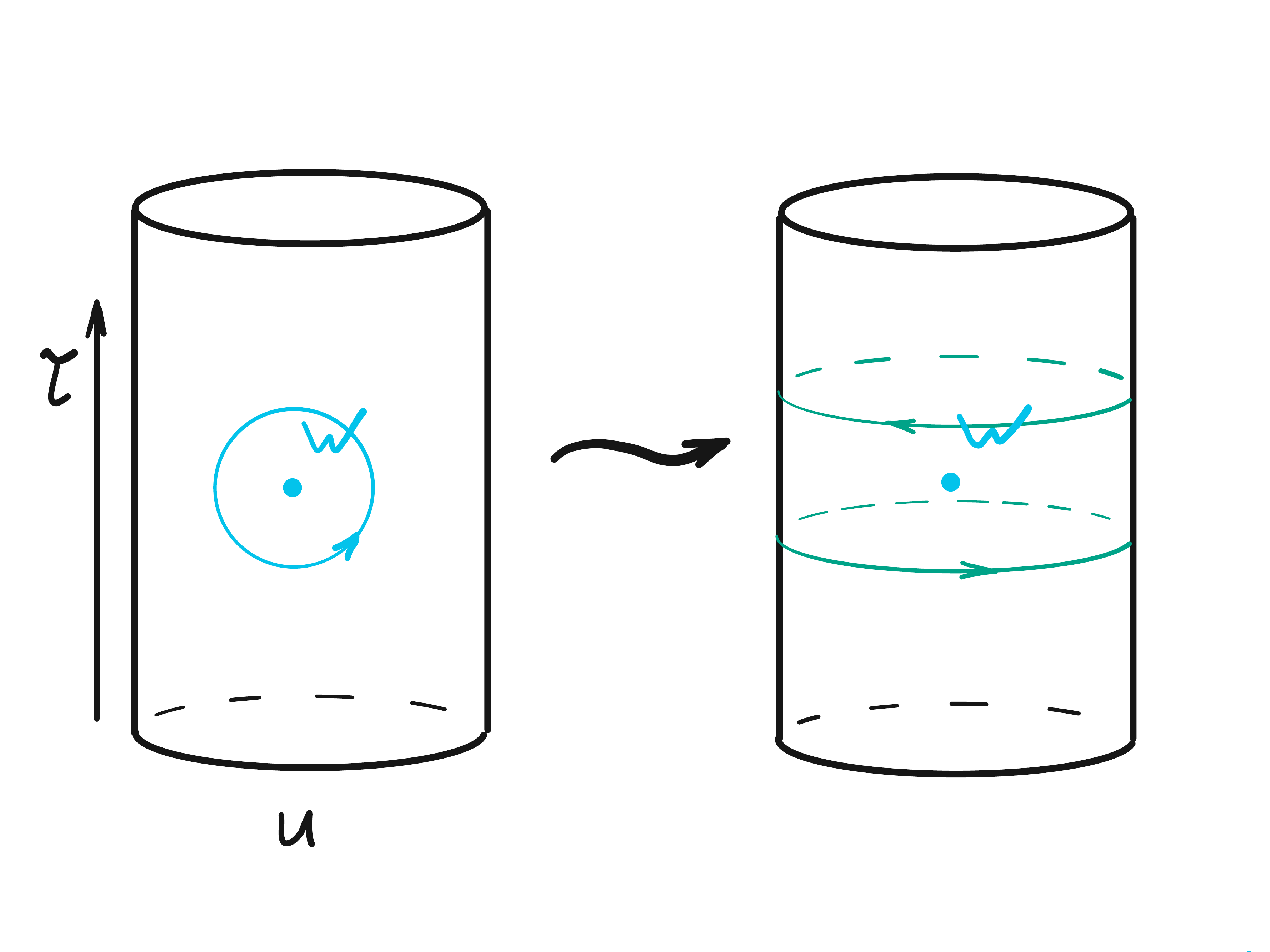}
\caption{\label{fig:i} Deformation of the blue contour (\ref{normal}) into the green one (\ref{cont2}).}
\end{figure}

To perform the integration (\ref{normal}) we will split the contour into two  pieces as showed in the fugure \ref{fig:i}, where we deform the contour in such a way that we can deal with two chronologically ordered expressions separately\footnote{We thank Pavlo Gavrilenko for help with the following calculation.}, namely
\bea
\label{cont2}
(AB)(w) = \int_{-i \epsilon}^{2\pi - i \epsilon} \frac{dz}{2\pi i} \frac{A(z)B(w)}{z-w} - \int_{i \epsilon}^{2\pi + i \epsilon} \frac{dz}{2\pi i} \frac{B(w)A(z)}{z-w}. 
\eea
The strategy of calculating (\ref{cont2}) is to express the integrand in terms of mode expansion $e^{iw}$ and $e^{iz}$. Let us introduce two axillary variables $u$ and $v$,
\bea
u = e^{iz}, \ \ v = e^{iw},
\eea
and make the following transformation,
\bea
\label{uv}
\begin{split}
\frac{1}{2 \pi i} \frac{dz}{z - w} = \frac{1}{2 \pi} \frac{dz}{\log(u) - \log(w)} = \frac{1}{2\pi} \frac{dz}{\log \left( 1 + \frac{u - v}{v}\right)} = \\ = \frac{dz}{2\pi} \left( \frac{v}{u -v} + \sum_{k = 0}^{\infty} c_k \left( \frac{u - v}{v} \right)^k  \right),
\end{split}
\eea
where we formally expanded one over logarithm and $c_k$ denote the coefficients of this expansion. The combination $v/(u-v)$ can be easily represented in terms of the mode expansion, namely 
\bea
\label{moderep}
\begin{split}
\frac{v}{u - v} =\left\{\begin{array}{c}
\sum_{k = 1}^{+\infty} e^{ik(w-z)}, \ \ \Im(z) < \Im(w), \\[8pt]
\-\sum_{k = 0}^{+\infty} e^{ik(z-w)}, \ \ \Im(z) > \Im(w).
\end{array} \right.
\end{split}
\eea
Plugging (\ref{uv}) and  (\ref{moderep}) into (\ref{cont2}) we obtain
\bea
\label{inter}
\begin{split}
(AB)(w) = \int_{-i \epsilon}^{2\pi - i \epsilon} \frac{dz}{2\pi} \left( \sum_{k =1}^{\infty} e^{ik(w -z)} + \sum_{k = 0}^{\infty} c_k\left(e^{i(z-w)} - 1\right)^k  \right) A(z)B(w) - \\ - \int_{+ i \epsilon}^{2\pi + i \epsilon} \frac{dz}{2\pi} \left( -\sum_{k=0}^{\infty} e^{ik(z-w)} + \sum_{k = 0}^{\infty} c_k\left(e^{i(z-w)} - 1\right)^k  \right)B(w) A(z).
\end{split}
\eea
In each of the two integrals in (\ref{inter}) we can integrate the first term assuming (\ref{modes}) and combine the rest into an integral of the commutator $[A(z), B(w)]$, i.e.
\bea
\label{nomralf}
\begin{split}
(AB)(w) = A_{-}(w) B(w) + B(w) A_{+}(w) + \\ + \int_{0}^{2\pi} \frac{dz}{2 \pi} \sum_{k = 0}^{\infty} c_k \left(e^{i(z-w)} - 1 \right)^k [A(z), B(w)],
\end{split}
\eea
where
\bea
A_{-}(w) = \sum_{n=1}^{\infty} A_{-n} e^{inw},  \ \ A_{+}(w) = \sum_{n=0}^{\infty} A_{n} e^{-inw}.
\eea
Substituting (\ref{nomralf}) into (\ref{zerom}) we obtain
\bea
\label{zerof}
\begin{split}
(AB)_0 = \sum_{n =1}^{\infty} A_{-n} B_n + \sum_{n=0}^{\infty} B_{-n} A_n + \\ + \int_{0}^{2 \pi} \frac{dw}{2\pi} \int_{0}^{2\pi} \frac{dz}{2 \pi} \sum_{k = 0}^{\infty} c_k \left(e^{i(z-w)} - 1 \right)^k [A(z), B(w)].
\end{split}
\eea
To move further one should calculate the commutator $[A(z), B(w)]$ and  perform the  integration of an infinite sum over $k$. This seems to be a difficult task, but we will see shortly that only finite number of terms  in this sum give non-zero contributions to the integral  
\bea
\int_{0}^{2\pi} \frac{dz}{2 \pi} \sum_{k = 0}^{\infty} c_k \left(e^{i(z-w)} - 1 \right)^k [A(z), B(w)]
\eea
from 
(\ref{nomralf}). Indeed,  if we fix $k$ and expand $\left(e^{i(z-w)}-1\right)^k$ we immediately see that the only modes of $A(z)$ which contribute are those from $0$ to $k$. More precisely for a given  fixed $k$ we have
\bea
\label{integral}
\sum_{n=0}^k c_k {k!\over n! (n-k)!} (-1)^n e^{-i n w} [A_n,B(w)] \label{Sti}
\eea
If we now rewrite the commutator in terms of the mode expansion $e^{-i n w} [A_n,B(w)]=\sum_m [A_n,B_m] e^{-(n+m)w}$ and keep in mind that both operators $A$ and $B$ are ``built'' out of stress-energy tensor, then both  $A_n$ and $B_n$ will be some normal ordered polynomials in $L_i$ such that the total sum of indexes is equal to $n$. 
The commutator $[A_n,B_m]$ is therefore also a polynomial in $L_i$ with a coefficient which is a polynomial in $n$. Therefore the sum in \eqref{Sti} will be a linear combination of the terms
\bea
S(a,k)=\sum_{n=0}^k {1\over n! (n-k)!} (-1)^{k-n} n^a, \label{Stirling}
\eea
where $a$ is some non-zero integer. The expression above is the Stirling number of the second kind, which vanishes unless $k\leq a$. This  immediately confirms that only finite number of terms in \eqref{integral} with $k\leq a$ contribute. 

We illustrate the emergence of the polynomial expression in $n$ explicitly in the case $A=B=T(w)$ in the next section. Here we only note that the finite number of terms contributing in the sum over $k$ in \eqref{nomralf} provide a crucial simplification. It allows computing  \eqref{nomralf} efficiently using computer algebra, which we use extensively in the computation of the zero modes in section \ref{sec:dim8} as well as of $Q_7$ in section \ref{sec:q7}.

\section{Warm-up: computation of $Q_3$}
\label{sec:q3}
In this section we apply the machinery devised in the previous section to the simplest non-trivial example $Q_3 = \oint \frac{dz}{2\pi} (TT)(z)$ and show explicitly how to perform the integration (\ref{zerof}). It's convenient to introduce shifted Virasoro generators,

\bea
\tilde{L}_n = L_n - \frac{c}{24} \delta_{n, 0}.
\eea
In terms of these operators the stress-energy tensor $T(z)$ has the following mode expansion
\bea
T(z) = \sum_{n = -\infty}^{+\infty} \tilde{L}_n e^{-inz},
\eea
and the Virasoro algebra in terms of the shifted generators is modified as
\bea
[\tilde{L}_n, \tilde{L}_m] = (n-m) \tilde{L}_{n+m} + \frac{c}{12} n^3 \delta_{n+m, 0}.
\eea
Substituting $A(z) = T(z)$ and $B(w) = T(w)$ into (\ref{zerof}) we obtain
 \bea
 \label{TTc}
 \begin{split}
 Q_3 = (TT)_0 = 2 \sum_{n =1}^{+\infty} \tilde{L}_{-n} \tilde{L}_n + \tilde{L}_0^2 + \\ + \int_{0}^{2 \pi} \frac{dw}{2\pi} \int_{0}^{2\pi} \frac{dz}{2 \pi} \sum_{k = 0}^{\infty} c_k \left(e^{i(z-w)} - 1 \right)^k [T(z), T(w)]. 
 \end{split}
 \eea
 
 The calculation of the commutator is a bit tedious but straightforward:
 \bea
 \begin{split}
 [T(z), T(w)] = \sum_{m, n \in \mathbb{Z}} e^{-inz} e^{-imw} [\tilde{L}_n, \tilde{L}_m] = \\ = \sum_{m, n \in \mathbb{Z}} e^{-imz} e^{-inw} \left( (n-m) \tilde{L}_{n+m}  + \frac{c}{12} n^3 \delta_{n+m, 0} \right) = \\ =  \sum_{m, n \in \mathbb{Z}} e^{-in(z-w)} (2n - (m+n)) e^{-i(m+n)w} \tilde{L}_{m+n} + \frac{c}{12} \sum_{n  \in \mathbb{Z}} n^3 e^{-in(z-w)}.
\label{p1}
 \end{split}
\eea
As was anticipated in the end of previous section, the commutator of two local operators gives rise to a polynomial  expression in $n$. Namely, using the notations of the previous section ($A_n$ is equal to $\tilde{L}_n$ and $B(w)$ is equal to $T(w)$),
\bea
e^{-i n w} [A_n,B(w)]=\sum_{m \in \mathbb{Z}}  (2n - m) e^{-i m w} \tilde{L}_{m} + \frac{c}{12} n^3, \nonumber
\eea
and therefore only terms with $k\leq 3$ will contribute  in the sum over $k$ in  \eqref{nomralf}. Here we would like to illustrate that by combing \eqref{p1} into a local expression. We continue,
\bea
\begin{split}
 [T(z), T(w)] =\\  2 (i \partial_z)  \sum_{n  \in \mathbb{Z}} e^{-in(z-w)} T(w) + i  \sum_{n  \in \mathbb{Z}} e^{-in (z-w)} \partial_w T(w)  + \frac{c}{12} (i \partial_z)^3 \sum_{n  \in \mathbb{Z}} e^{-in(z-w)} = \\ =4 \pi i \partial_z \delta(z - w) T(w) + 2 \pi i \delta(z-w) \partial_w T(w) +  \frac{c}{12} (i \partial_z)^3 2\pi \delta(z - w),
 \end{split}
 \eea
where on the last line we have used the following representation of delta-function, 
\bea
2 \pi \delta(z -w) = \sum_{n  \in \mathbb{Z}} e^{-in (z-w)} .
\eea
Note, that we have represented the commutator of stress-energy tensors $[T(z), T(w)]$ in such a way that every term in the final expression contains delta-function $\delta(z - w)$ or its derivatives. This representation helps us easily perform the integration over $z$ in (\ref{TTc}), namely

\bea
\begin{split}
\label{TTint}
\oint \frac{dz}{2 \pi} \sum_{k = 0}^{\infty} c_k \left(e^{i(z-w)} - 1 \right)^k [T(z), T(w)]  =  i c_0 \partial_w T(w)  -  \\   \left. -2 i \partial_z \sum_{k = 0}^{+\infty} c_k \left(e^{i(z-w)} - 1 \right)^k \right \vert_{z = w} T(w)+\frac{c}{12}  \left. (-i \partial_z)^3 \sum_{k = 0}^{+\infty} c_k \left(e^{i(z-w) - 1} \right)^k \right \vert_{z = w}. 
\end{split}
\eea
Here in the first term we have used $0^0 = 1$. Let us denote
\bea
a_k  = \left. -i \partial_z \left(e^{i(z-w)} - 1 \right)^k  \right \vert_{z = w},  \ \ b_k = \left.    (-i \partial_z)^3  \left(e^{i(z-w)} - 1 \right)^k \right  \vert_{z = w}.
\eea
The coefficients $a_k$ and $b_k$ are non-zero only for the  first few values of $k\leq 3$.  Specifically, 
\bea
\begin{split}
a_1 = 1,  \ \ a_{2, 3, 4, ...} = 0, \\
b_1 =1,  \ \ b_2 = 6,  \ \ b_3  = 6,  \ \ b_{4, 5, 6, ...} = 0.
\end{split}
\eea

That means that the sums $\sum c_k\, a_k$ and $\sum c_k\, b_k$ contain only finite number of terms and only four first coefficients $c_k$ contribute to these sums. From the definition we find
\bea
c_0 = \frac{1}{2}, \ \ c_1 = - \frac{1}{12},  \ \ c_2 = \frac{1}{24}, \ \ c_3 = -\frac{19}{720}.
\eea
Combining (\ref{TTint}) and (\ref{nomralf}) we get the normal ordered expression 
\bea
(TT)(w) = T_-(w) T(w) + T T_+(w) - \frac{1}{6} T(w) + \frac{i}{2} \partial_w T(w) + \frac{c}{1440},
\eea
where $T_-(w) = \sum_{k=1} e^{ikw} \tilde{L}_{-k} $ and $T_+(w) = \sum_{k=0} e^{-ikw} \tilde{L}_{k} $.

And, finally, for zero mode
\bea
\label{q3vir}
\begin{split}
Q_3 = (TT)_0 = \oint \frac{dw}{2 \pi} (TT)(w) = 2 \sum_{n = 1}^{+\infty} \tilde{L}_{-n} \tilde{L}_n + \tilde{L}_0^2 - \frac{1}{6} \tilde{L}_0 + \frac{c}{1440} = \\ = 2 \sum_{n = 1}^{+\infty} L_{-n} L_n + L_0^2 - \frac{c+2}{12} L_0 + \frac{c(5c + 22)}{2880}.
\end{split}
\eea
The result matches that one of  \cite{bazhanov1996integrable}. The technique we have described here in principle can be applied to a calculation of zero modes of any product of local operators.

\section{Quasi-primaries}
\label{sec:QP}
In this section we list the  explicit expressions for the zero modes of all quasi-primaries 
from the vacuum family with the dimension less or equal to eight. Up to dimension nine all quasi-primaries have even dimension. There is a unique operator of dimension zero -- the identity operator, which is a primary. There is also a unique quasi-primary at level two, the stress-energy tensor $T$. Its zero mode is the CFT Hamiltonian -- first KdV charge  $Q_1=L_0-c/24$.
At level two there is also a unique quasi-primary
\bea
T_2=(TT)-{3\over 10}\partial^2 T.
\eea 
Its zero mode is the first non-trivial KdV charge $Q_3$ given by \eqref{q3vir}.
At all other levels the quasi-primaries are not unique and we  organize them by dimension, nested powers of $T$ and orthogonality of Zamolodchikov metric. 

\subsection{Quasi-primaries of dimension 6}
At level four there are two quasi-primaries, 
\bea
\mathcal{B}&=&(\partial T \partial T) - \frac{4}{5} (\partial^2 T T) - \frac{1}{42} \partial^4 T,
\eea
and 
\bea
\mathcal{D} = (T(TT)) - \frac{9}{10} (\partial^2 T T) - \frac{1}{28} \partial^4 T + \frac{93}{70c + 29} \mathcal{B}.
\eea
Zero mode of their combination $\mathcal{D} - \frac{5(43 + 14c)}{2(29 + 70c)} \mathcal{B}$ is the KdV charge $Q_5$, which was found explicitly in \cite{bazhanov1996integrable}. To find the explicit form of $\mathcal{B}_0$ and $\mathcal{D}_0$ we introduce the ``building block''
\bea
(\partial T \partial T)_0 =- (\partial^2 T T)_0 =  2 \sum_{n=1}^{\infty} n^2 \tilde{L}_{-n}  \tilde{L}_n + \frac{\tilde{L}_0}{60} - \frac{c}{3024}, 
\eea
which is different from the quasi-primary $\mathcal{B}$ by a total derivative. Therefore 
\bea
\mathcal{B}_0&=& \frac{9}{5} (\partial T \partial T)_0.
\eea
Similarly, to calculate $\mathcal{D}_0$ we introduce 
\bea
\begin{split}
(T(TT))_0 = \sum_{k, l = 0}^{\infty} \tilde{L}_{-k-l} \tilde{L}_k \tilde{L}_l + 2 \sum_{k = 1, l = 0}^{\infty} \tilde{L}_{-k} \tilde{L}_{k-l} \tilde{L}_l + \sum_{k, l =1}^{\infty} \tilde{L}_{-k} \tilde{L}_{-l} \tilde{L}_{k+l} - \\ - \sum_{n=1}^{\infty} \tilde{L}_{-n} \tilde{L}_n - \frac{\tilde{L}_0^2}{2} + \frac{c \tilde{L}_0}{480} + \frac{\tilde{L}_0}{15} - \frac{c}{3024},
\end{split}
\eea
such that 
\bea
\mathcal{D}_0 = (T(TT))_0 + \frac{9}{10} (\partial T \partial T)_0.
\eea

\subsection{Quasi-primaries of dimension 8}
\label{sec:dim8}
At level eight there are three quasi-primaries,
\bea
\mathcal{E} = (\partial^2 T \partial^2 T) - \frac{10}{9} (\partial^3 T \partial T) + \frac{10}{63} (\partial^4 T T) - \frac{1}{324} \partial^6 T,
\eea
\bea
\mathcal{H} = (\partial T (\partial T T)) - \frac{4}{5} (\partial^2 T (TT)) + \frac{2}{15} (\partial^3 T \partial T) - \frac{3}{70} (\partial^4 T T) + \frac{9(140c + 83)}{50(105c +11)} \mathcal{E},
\eea
and
\bea
\begin{split}
\mathcal{I} = (T(T(TT))) - \frac{9}{5} (\partial^2 T (TT)) + \frac{3}{10} (\partial^3 T \partial T) + \frac{81(35c - 51)}{100(105c + 11)} \mathcal{E} + \\+ \frac{12(465c -127)}{5c(210c + 661) - 251} \mathcal{H}.
\end{split}
\eea

The ``building block'' for calculating $\mathcal{E}_0$ is
\bea
(\partial^2 T \partial^2 T)_0 = - (\partial^3 T \partial T)_0 = (\partial^4 T  T)_0 = 2 \sum_{n=1}^{\infty} n^4 \tilde{L}_{-n} \tilde{L}_n - \frac{\tilde{L}_0}{126} + \frac{c}{2880},
\eea
And therefore, 
\bea
\mathcal{E}_0 = \frac{143}{63} (\partial^2 T \partial^2 T)_0.
\eea

There are two ``building blocks'' for $\mathcal{H}$:
\bea
\begin{split}
(\partial T (\partial T T))_0 = - \sum_{k = 0, l = 0}^{\infty} kl \tilde{L}_{-k-l} \tilde{L}_k \tilde{L}_l + 2 \sum_{k = 1, l = 0}^{\infty} k l \tilde{L}_{-k} \tilde{L}_{k-l} \tilde{L}_l - \\ - \sum_{k, l = 1}^{\infty} kl \tilde{L}_{-k} \tilde{L}_{-l} \tilde{L}_{k+l} + \frac{1}{12}(\partial T \partial T)_0 + \frac{1}{30} \sum_{n=1}^{\infty} \tilde{L}_{-n} \tilde{L}_n + \frac{\tilde{L}_0^2}{60} -\frac{(5 c+93) \tilde{L}_0}{15120} + \frac{113c}{1814400}
\end{split}
\eea
and 
\bea
\begin{split}
(\partial^2 T ( T T))_0 = - \sum_{k, l = 0}^{\infty} l^2 \tilde{L}_{-k-l} \tilde{L}_k \tilde{L}_l - \sum_{k=1, l=0}^{\infty} (k^2 + l^2) \tilde{L}_{-k} \tilde{L}_{k-l} \tilde{L}_l - \\ - \sum_{k=1, l=1}^{\infty} k^2 \tilde{L}_{-k} \tilde{L}_{-l} \tilde{L}_{k+l} + \frac{7}{120} \sum_{n= 1}^{\infty} \tilde{L}_{-n} \tilde{L}_n - \frac{\tilde{L}_0^2}{30} + \frac{c \tilde{L}_0}{1512} + \frac{61\tilde{L}_0}{7560} - \frac{131c}{1814400}.
\end{split}
\eea

The zero mode of $\mathcal{H}$ is
\bea
\mathcal{H}_0 = (\partial T(\partial T T))_0 -\frac{4}{5} (\partial^2 T (TT))_0 + \frac{7035 c+13652}{110250 c+11550} (\partial^2 T \partial^2 T)_0.
\eea

The expression for $(T(T(TT)))_0$ is too bulky to write it twice. We do not write it explicitly here, but simply mention that it can be obtained from the Virasoro algebra expression for $Q_7$, which we give explicitly in the next section, by subtracting $(\partial^2 T \partial^2 T)_0$ and $(T (\partial T \partial T))_0$ with proper coefficients, see equation \eqref{J8}. Thus, in lieu  of  $(T(T(TT)))_0$ we give explicitly the expression for $(T (\partial T \partial T))_0$,
\bea
\begin{split}
(T (\partial T \partial T))_0 = \sum_{k, l = 1}^{\infty} (k+l) l \tilde{L}_{-k} \tilde{L}_{-l} \tilde{L}_{k+l} + \sum_{l=0, k=1}^{\infty}(k-l)k \tilde{L}_{-k} \tilde{L}_{k-l} \tilde{L}_l + \\ + \sum_{k, l = 0}^{\infty} (k+l) k \tilde{L}_{-k-l} \tilde{L}_k \tilde{L}_l + \sum_{k=0, l=1}^{\infty} (k-l)k \tilde{L}_{-l} \tilde{L}_{l-k} \tilde{L}_k - \\ - \frac{7}{6} \sum_{n=1}^{\infty} n^2 \tilde{L}_{-n} \tilde{L}_n + \frac{1}{30} \sum_{n=1}^{\infty} \tilde{L}_{-n} \tilde{L}_n + \frac{\tilde{L}_0^2}{60} + \frac{c \tilde{L}_0}{3024} - \frac{\tilde{L}_0}{135} + \frac{13c}{86400}.
\end{split}
\eea
Finally, for the last quasi-primary at level 8 we have
\bea
\begin{split}
\mathcal{I}_0 = (T(T(TT)))_0 - \frac{9}{5} (\partial^2 T (TT))_0 - \frac{3}{10} (\partial^2 T \partial^2 T)_0 + \frac{81(35c - 51)}{100(105c + 11)} \mathcal{E}_0 + \\+ \frac{12(465c -127)}{5c(210c + 661) - 251} \mathcal{H}_0.
\end{split}
\eea

\section{Expression for $Q_7$}
\label{sec:q7}
In this section we present the expression for $Q_7$ in terms of Virasoro generators, which we calculated by applying the technique described above. Then we test our result by showing that it is consistent with the known spectrum $Q_7$ at the leading $1/c$ order. We further check at that our expression for $Q_7$ vanishes as an operator for the $(9, 2)$ minimal model at the first dozen of descendant levels, as is predicted in \cite{maloney2018thermal,bazhanov1996integrable}. 
Finally, we will show how to use commutativity of qKdV charges, known results about spectrum in $1/c$ expansion and the constraints from the minimal models to get a shortcut for the expression \eqref{q7}.

\subsection{The result}
$Q_7$ is the zero mode of the operator
\bea
\label{J8}
J_8 = (T(T(TT))) + \frac{c+2}{3}(T(\partial T \partial T)) + \frac{2c^2 -17c - 42}{360} (\partial^2 T \partial^2 T).
\eea
In principle $J_8$ (and higher densities $J_{2n}$) can be determined by requiring commutativity of $Q_7$ with $Q_5$ and $Q_3$, however, the calculation is quite involved. Alternatively, one can use the expression for the thermal correlation function $\langle Q_{2n-1} \rangle$, which can be calculated using other means and fix the coefficients of $J_{2n}$ in  this way. This was  done for $J_8$, $J_{10}$ and $J_{12}$ in \cite{maloney2018thermal}. However, for $J_{14}$ and higher densities the number of independent coefficients becomes too large to be uniquely fixed from the form of $\langle Q_{2n-1} \rangle$ alone.

We find the following expression for $Q_7$ in terms of the Virasoro algebra generators 
\bea
\label{q7}
\begin{split}
Q_7 = \sum_{k, l, m = 1}^{\infty} L_{-k} L_{-l} L_{-m} L_{k+l+m} + \sum_{k, l, m = 0}^{\infty} L_{-k-l-m} L_{k} L_{l} L_{m} + \\ + 3\sum_{\substack{k, l = 1  \\ m= 0}}^{\infty} L_{-k} L_{-l} L_{k+l-m} L_m + 3 \sum_{\substack{k = 1  \\ l, m= 0}}^{\infty} L_{-k} L_{k-l-m} L_l L_m + \\ \frac{8+c}{3} \left[\sum_{k, l =1}^{\infty} (k+l)l L_{-k} L_{-l}L_{k+l} + \sum_{\substack{k = 1  \\ l= 0}}^{\infty} (k-l)k L_{-k} L_{k-l} L_l \right] + \\ + \frac{8+c}{3} \left[\sum_{k, l =0}^{\infty} (k+l)k L_{-k-l} L_{k}L_{l} + \sum_{\substack{k = 0  \\ l= 1}}^{\infty} (k-l)k L_{-l} L_{l-k} L_k \right] + \\ + \sum_{n=1}^{\infty} \left(\frac{c^2 - c -141}{90}n^4 - \frac{7c + 59}{18}n^2 \right) L_{-n}L_n - \left(\frac{1}{48}c^2 +\frac{53}{360} c + \frac{19}{90} \right) \tilde{Q}_3 - \\ - \left(\frac{1}{6} c + 1 \right) \tilde{Q}_5  - \frac{c+6}{6} L^3_0 + \frac{15c^2 + 194c + 568}{1440}L_0^2 - \\ -\frac{(c+2)(c+10)(3c+28)}{10368} L_0  + \frac{c(3c+46)(25c^2+ 426c + 1400)}{24883200}.
\end{split}
\eea
This is one of the main results of this paper. 
Here $\tilde{Q}_3$ and $\tilde{Q}_5$ are defined as parts of  $Q_3$ and $Q_5$ which annihilate the primary states,  $\tilde{Q}_3 \ket{\Delta} = 0$ and  $\tilde{Q}_5 \ket{\Delta} = 0$, namely, 
\bea
\tilde{Q}_3 = 2 \sum L_{-n}L_n,
\eea
and 
\bea
\begin{split}
\label{Q5t}
\tilde{Q}_5 = \sum_{k, l = 0}^{\infty} L_{-k-l} L_k L_l + 2\sum_{k = 1, l = 0}^{\infty} L_{-k} L_{k-l} L_l + \sum_{k, l = 1}^{\infty} L_{-k} L_{-l} L_{k+l} \, + \\ + \sum_{n=1}^{\infty} \left(\frac{c+2}{6}n^2 -\frac{c}{4} - 1 \right) L_{-n}L_n - L_0^3.
\end{split}
\eea
The expression (\ref{Q5t}) can also be represented in a slightly different way \cite{bazhanov1996integrable} using the following identity,
\bea
\begin{split}
\sum_{k, l = 0}^{\infty} L_{-k-l} L_k L_l + 2\sum_{k = 1, l = 0}^{\infty} L_{-k} L_{k-l} L_l + \sum_{k, l = 1}^{\infty} L_{-k} L_{-l} L_{k+l} = \\  \sum_{n_1 + n_2 + n_3 = 0} :L_{n_1} L_{n_2} L_{n_3}: + \frac{3}{2} \sum_{n=1} n^2 L_{-n} L_n + \frac{3}{2} \sum_{r = 1} L_{1-2r} L_{2r-1}.
\end{split}
\eea

\subsection{The consistency check: $1/c$ expansion and the (9, 2) minimal model}
In this section we perform two different consistency checks of the Virasoro algebra  expression for $Q_7$ \eqref{q7}.

In an upcoming paper  \cite{upcoming} we have calculated the spectrum of all KdV charges using semi-classical quantization at the first few orders in $1/c$ expansion. This calculation did not rely on the explicit form of $Q_{2k-1}$ in terms of the Virasoro algebra generators, and hence can be used to cross-check our result. 

The spectrum in $1/c$ expansion can be parametrized by ``quantum numbers'' of the conventional basis  in the space of descendants, namely
\bea
\ket{\{ m_i\}, \Delta}  = L_{-m_1} ...L_{-m_k} \ket{\Delta}.
\eea
Each set $\{ m_i \}$ can be rewritten using the so-called boson representation. Namely, each set $\{ m_i \}$ can be parametrized by the set of integers $\{ n_k \}$, where $n_k$ is the number of times natural number $k$ appears in $\{ m_i \}$. Therefore, for any integer $p$,
\bea
\sum_i m_i^p = \sum_{k = 1}^{\infty} k^p n_k,
\eea
Leading $1/c$  spectrum of $Q_7$ in these notations is given by $Q_7 \ket{\lambda} = \lambda \ket{\lambda}$, where 
\bea
\begin{split}
\lambda = \Delta'^4 + \Delta'^3 \left( 28 \sum_k n_k k - 1 \right) + \Delta'^2 c \left(\frac{7}{3} \sum_k n_k k^3 + \frac{7}{720}  \right) +\\ + \Delta' c^2 \left( \frac{7}{90} \sum_k n_k k^5 - \frac{1}{6480} \right) + c^3 \left(\frac{1}{1080} \sum_k n_k k^7 + \frac{1}{518400} \right) + \\  + \Delta'^2 \left( 98 n^2 - 77 \sum_k n_k^2 k^2  + \frac{259}{3} \sum_k n_k k^3 -77 \sum_k n_k k^2 - \frac{14}{3} \sum_k n_k k + \frac{71}{180}\right) +\\+ \Delta' c \left( \frac{98}{15}   
\sum_{k, l} k^3 l n_k n_l - \frac{56}{15} \sum_k n_k^2 k^4 + \frac{63}{25} \sum_k n_k k^5 - \right. \\ \left.-\frac{56}{15} \sum_k n_k k^4 - \frac{7}{90} \sum_k n_k k^3 + \frac{49}{1800} \sum_k n_k k - \frac{23}{4320}  \right) + \\+ c^2 \left(\frac{7}{180} \sum_{k, l} n_k n_l k^3 l^3 + \frac{7}{90} \sum_{k, l}n_k n_l k^5 l - \frac{7}{120} \sum_k n_k^2 k^6 + \frac{127}{5400} \sum n_k k^7 \right. \\\left. -\frac{7}{120} \sum_k n_k k^6 + \frac{7}{21600} \sum_k n_k k^3 - \frac{1}{6480} \sum_k n_k k + \frac{103}{2073600} \right) - \\ - 504\frac{\Delta'^3}{c} \left(\sum_k n_k^2 - 2\sum_k  n_k k + \sum_k n_k \right) + O(c)
\end{split}
\label{Q7sp}
\eea
Using computer algebra one can check explicitly at first dozen of descendant levels that this spectrum matches with the one following from (\ref{q7}).

Another check is provided by \cite{maloney2018thermal,bazhanov1996integrable}, which shows that for the minimal models $(2n+3, 2)$, $n \leqslant 1$,  the qKdV charges $Q_k$ with $k$ divisible by $2n+1$ vanish as operators. Hence $Q_7$ should vanish as an operator in the minimal model $(9, 2)$, with $n = 3$ and central charge  $c = -46/3$. This minimal model includes primaries with the following  dimensions  $\{ \Delta_k\} = \{ 0, -1/3, -2/3, -5/9\}$. Using computer algebra we have verified that $Q_7$ vanishes for all non-zero states of this model up to the descendant level twelve.

\subsection{A shortcut}
In this subsection we show how to get the Virasoro algebra expression for $Q_7$ without full explicit calculation of all   involved  commutators  by exploiting the restrictions from the commutativity of qKdV charges, $1/c$ expansion and the $(9, 2)$ minimal model. Our goal here will be to understand what kind of terms may appear  in the Virasoro algebra expression of $Q_7$ and then fix the coefficients using the  constraints. 

We start from the last term of $J_8$ (\ref{J8}):
\bea
A_8 = (\partial^2 T \partial^2 T).
\eea
Substiting $A_{8}$ into (\ref{nomralf} - \ref{zerof}) we obtain
\bea
\oint \frac{dw}{2 \pi} A_8 \sim \sum_{n=1}^{\infty} n^4 L_{-n} L_n + comm,
\eea
where $comm$ comes from the integral of the commutator in (\ref{nomralf}). Expressing the commutator in terms of Virasoro generators we get $[\partial^2 T, \partial^2 T] \sim [L_n, L_m] 
\sim L_k$ due to Virasoro algebra. It is easy to see that the final answer for the zero mode should contain only such operators that map states of level $k$ to the states of level $k$. That means that after integrating the commutator we can only get some function of $L_0$. Therefore, 
\bea
\oint \frac{dw}{2 \pi} A_8 \sim \sum_{n=1}^{\infty} n^4 L_{-n} L_n + f(L_0).
\eea
Now we turn to the next term 
\bea
B_8 = (T(\partial T \partial T)).
\eea
The term $B_8$ contains nested normal ordering. We will deal with it subsequently,
\bea
\label{dTdT}
(\partial T \partial T)(w) = \partial T \partial T_{+} + \partial T_{-} \partial T  + \oint \frac{dz}{2\pi} \sum_{k=0}^{\infty} c_k \left( e^{i(z-w)} - 1\right)^k [\partial T(z), \partial T(w)].
\eea
One can calculate the commutator and perform the integration (\ref{dTdT})  explicitly but we just notice that the result of the integration can only contain terms with one stress-energy tensor $T$ or its derivatives and the terms like $TT$ or $T \partial T$ are not present due to Virasoro algebra, which means
\bea
(\partial T \partial T)(w) = \partial T(w) \partial T_{+}(w) + \partial T_{-}(w) \partial T(w)  + f(T(w), \partial T(w), \partial^2 T(w)),
\eea
where $f(T, \partial T, \partial^2 T)$ is some linear function of its arguments. Substituting (\ref{dTdT}) into $B_8$ we get
\bea
\label{TdTdT}
(T(\partial T \partial T)) = T_{-} \partial T \partial T_{+} + \partial T \partial T_{+} T_{+} + T_{-} \partial T_{-} \partial T  + \partial T_{-} \partial T T_{+} + comm.
\eea
Here $comm$ again denotes some expression associated with the commutators, which we will not calculate explicitly. But let us notice again that it contains at most two stress-tensors or its derivatives. The only expression that survives after integration of such terms is proportional to $\sim L_{-n} L_n$, namely
\bea 
\oint dz \partial^{\alpha} T \partial^{\beta} T \sim \sum_{-\infty}^{+\infty} n^{\alpha + \beta} L_{-n} L_n.
\eea
Counting the amount of dervatives and integrating (\ref{TdTdT}) we obtain
\bea
\begin{split}
\oint \frac{dw}{2 \pi} B_8 = \sum_{k, l =1}^{\infty} (k+l)l L_{-k} L_{-l}L_{k+l} + \sum_{\substack{k = 1  \\ l= 0}}^{\infty} (k-l)k L_{-k} L_{k-l} L_l  + \\ +  \sum_{k, l =0}^{\infty} (k+l)k L_{-k-l} L_{k}L_{l} + \sum_{\substack{k = 0  \\ l= 1}}^{\infty} (k-l)k L_{-l} L_{l-k} L_k  + \\ + \sum_{n=1}^{\infty} \left( \alpha n^4 + \beta n^2 + \gamma  \right) L_{-n} L_n + f(L_0),
\end{split}
\eea
where $\alpha$, $\beta$, $\gamma$ are some $c$-dependent coefficients and $f(L_0)$ is some function of $c$ and $L_0$.

In the same manner one can deal with the term $(T(T(TT))$. As a result we get an expression for $Q_7$ with some coefficients to fix. The expression is 
\bea
\label{q7}
\begin{split}
Q_7 = \sum_{k, l, m = 1}^{\infty} L_{-k} L_{-l} L_{-m} L_{k+l+m} + \sum_{k, l, m = 0}^{\infty} L_{-k-l-m} L_{k} L_{l} L_{m} + \\ + 3\sum_{\substack{k, l = 1  \\ m= 0}}^{\infty} L_{-k} L_{-l} L_{k+l-m} L_m + 3 \sum_{\substack{k = 1  \\ l, m= 0}}^{\infty} L_{-k} L_{k-l-m} L_l L_m + \\ \frac{8+c}{3} \left[\sum_{k, l =1}^{\infty} (k+l)l L_{-k} L_{-l}L_{k+l} + \sum_{\substack{k = 1  \\ l= 0}}^{\infty} (k-l)k L_{-k} L_{k-l} L_l \right] + \\ + \frac{8+c}{3} \left[\sum_{k, l =0}^{\infty} (k+l)k L_{-k-l} L_{k}L_{l} + \sum_{\substack{k = 0  \\ l= 1}}^{\infty} (k-l)k L_{-l} L_{l-k} L_k \right] + \\  + \sum_{n=1}^{\infty} \left( \alpha n^4 + \beta n^2 \right) L_{-n}L_n + \gamma \tilde{Q}_3  + \delta  \tilde{Q}_5+ f(L_0)
\end{split},
\eea
where $\alpha$, $\beta$, $\gamma$ and $\delta$ are the coefficients dependent on central charge and $f(L_0)$ is some polynomial of $L_0$ and central charge.

The term $f(L_0)$ determines the value of $Q_7$ on a primary state. That has been previously calculated in the appendix B of \cite{bazhanov97zero}. The coefficients $\alpha$ and $\beta$ can be found from the commutativity constraint $[Q_3, Q_7] = 0$. From $1/c$ expansion of the spectrum (\ref{Q7sp}) we see that $\delta$ is at most linear polynomial in $c$, $\delta = \delta_1 c + \delta_2$ and $\gamma$ is at most quadratic polynomial in $c$,  $\gamma = \gamma_1 c^2 + \gamma_2 c + \gamma_3$. The coefficients $\delta_1$, $\delta_2$ and $\gamma_1$, $\gamma_2$ can be extracted directly from (\ref{Q7sp}) and, finally, $\gamma_3$ can be fixed by requiring that $Q_7$ vanishes acting on any descendant state of the (9,2) minimal model.

\section{Conclusions}
\label{sec:con}
In this paper we pedagogically developed and presented  the machinery of calculating the zero modes of local operators in a 2d CFT on a cylinder. We focused on the situation when the operators are from the vacuum family, i.e.~are built from the powers of stress-energy tensor and its derivatives. We have calculated  explicit expressions in terms of the Virasoro algebra generators for all quasi-primaries with the dimension of less or equal than eight. We have also calculated the explicit expression for the KdV charge $Q_7$, thus bringing the number of explicitly known charges to three (excluding $Q_1$, which is trivial). The explicit formulae obtained in this paper can be used to study spectral properties of the qKdV hierarchy and further investigate Eigenstate Thermalization Hypothesis in 2d CFTs, in particular by means of computer algebra.

\acknowledgments
We would like to thank Mikhail Bershtein,  Ilya Vilkoviskiy  and especially Pavlo Gavrilenko for discussions. AD is supported by the National Science Foundation under Grant No.~PHY-1720374. KP is supported by the RFBR grant  19-32-90173.



\bibliographystyle{JHEP}
\bibliography{Q7}

%
%
%
%
%
%
%
%
\end{document}